# Uniform doping of graphene close to the charge neutrality point by polymer-assisted spontaneous assembly of molecular dopants

**Hans He[1], Kyung Ho Kim[1,2], Andrey Danilov[1], Domenico Montemurro[1], Liyang Yu[3], Yung Woo Park[2,4,5], Floriana Lombardi[1], Thilo Bauch[1], Kasper Moth-Poulsen[3] , Tihomir Iakimov[6], Rositsa Yakimova[6], Per Malmberg[3], Christian Müller[3], Sergey Kubatkin[1] & Samuel Lara-Avila[1,7]**

1. Department of Microtechnology and Nanoscience, Chalmers University of Technology, SE-412 96, Gothenburg, Sweden

2 Department of Physics and Astronomy, Seoul National University, Seoul, 08826, Korea

3 Department of Chemistry and Chemical Engineering, Chalmers University of Technology, 41296 Göteborg, Sweden

4 Institute of Applied Physics, Seoul National University, Seoul, 08826, Korea

5 Department of Physics and Astronomy, University of Pennsylvania, Philadelphia, PA, 19104, USA

6 Department of Physics, Chemistry and Biology, Linkoping University, S-581 83, Linköping, Sweden

7 National Physical Laboratory, Hampton Road, Teddington, TW11 0LW, UK

**Tuning the charge carrier density of two-dimensional (2D) materials by incorporating dopants into the crystal lattice is a challenging task[1,2]. An attractive alternative is the surface transfer doping by adsorption of molecules on 2D crystals, which can lead to ordered molecular arrays[3–7]. However, such systems, demonstrated in ultra-high vacuum conditions (UHV), are often unstable in ambient conditions. Here we show that air-stable doping of epitaxial graphene on SiC - achieved by spin-coating deposition of 2,3,5,6-tetrafluoro-tetracyano-quino-dimethane (F4TCNQ) incorporated in poly (methyl-methacrylate) - proceeds via the spontaneous accumulation of dopants at the graphene-polymer interface and by the formation of a charge-transfer complex that yields low-disorder, charge-neutral graphene with carrier mobilities ~70,000 cm$^2$/Vs at cryogenic temperatures. The assembly of dopants on 2D materials assisted by a polymer matrix, demonstrated by spin coating wafer-scale substrates in ambient conditions, opens up a scalable technological route towards expanding the functionality of 2D materials.**

High performance solid-state devices based on 2D materials require homogeneous charge carrier density distribution down to the microscopic scale. For instance, epitaxial graphene grown on SiC (SiC/G) has been processed to the, hitherto, most demanding electronic application: a quantum resistance standard to determine the von Klitzing constant $R_K=h/e^2$ with relative standard uncertainty of 86 part-per-trillion[8–10]. Such quantum Hall devices demand spatially homogeneous doping, high mobility of charge carriers, and temporal stability at ambient conditions. For SiC/G this is a challenge, because such graphene is known to have high intrinsic n-doping (n>$10^{13}$ cm$^{-2}$) due to interaction with the substrate, making it difficult to substantially tune its carrier concentration[11].

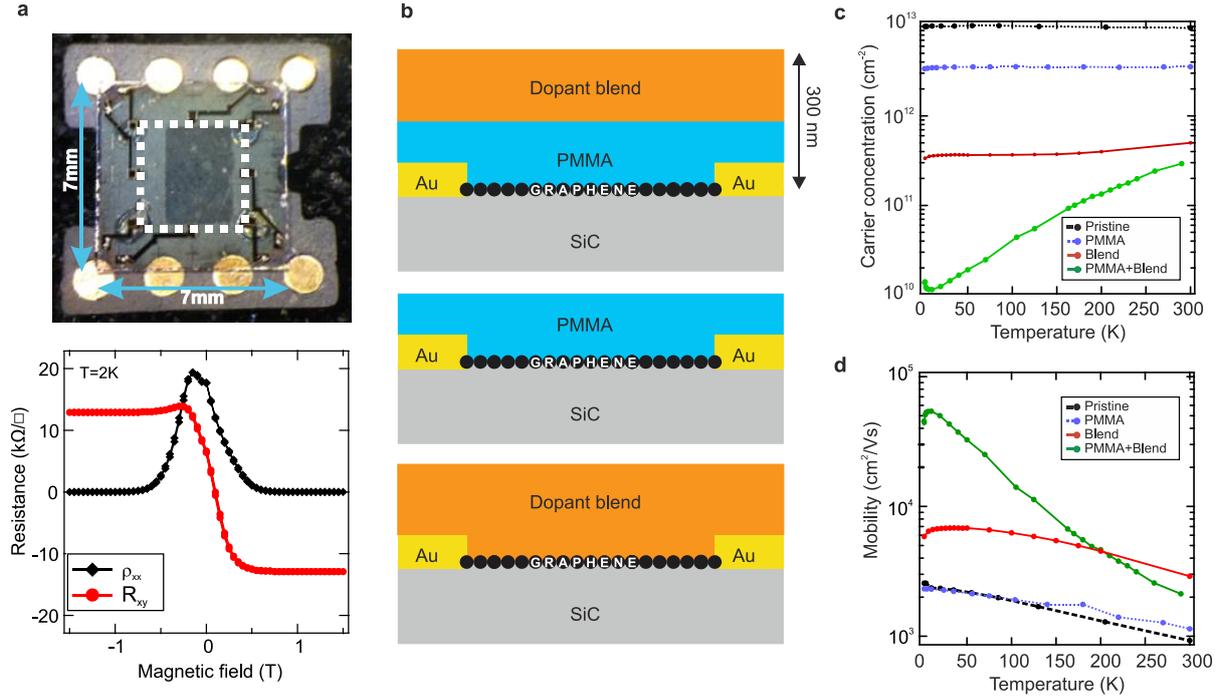

**Figure 1.** Magnetotransport characterization of chemically doped SiC/G. (a) Top: macroscopic graphene Hall bar device (white dotted outline, W=5 mm x L=5 mm). Bottom: Low field magnetoresistance and fully developed quantum Hall effect indicating low charge disorder in chemically doped graphene, even over macroscopic areas. For this device, the measured carrier density $p=9\times10^9$ cm$^{-2}$ and mobility $\mu= 39,000$ cm$^2$/Vs (b) The different encapsulation schemes with a polymer stack consisting of polymer-F4TCNQ dopant blend which comprises of F4TCNQ molecules in a PMMA matrix (F4TCNQ 7 wt.%). The three schemes are dopant blend separated from graphene by a PMMA spacer (top), only PMMA layer (middle), and the dopant blend directly on the surface of graphene (bottom). (c) Carrier density as a function of temperature extracted from Hall measurements on small epitaxial graphene devices (W=2-50 μm x L=10-100 μm). (d) The corresponding Hall carrier mobility showing the highest $\mu\sim55,000$ cm$^2$/Vs at T=10K for sample prepared with PMMA spacer and dopant layer. The downturn in mobility at lower temperatures is due to quantum corrections to Drude resistance. Carrier concentration n and mobilities μ were extracted from Hall measurements as $n=1/eR_H$ and $\mu= R_H/\rho_{XX}$, with e the elementary charge, the Hall coefficient $R_H=dR_{XY}/dB$, the longitudinal sheet resistance $\rho_{XX}=R_{XX}W/L$, and $R_{XY}$ the transversal resistance.

Here we show a molecular approach to dope epitaxial graphene homogeneously, yielding graphene with low carrier density (n<$10^{10}$ cm$^{-2}$), low charge fluctuations (at the level of $\Delta n\approx\pm6\times10^9$ cm$^{-2}$), and carrier mobilities up to 70,000 cm$^2$/Vs over macroscopic areas at low temperatures. So far, such electron transport properties have only been attained in microscopic single crystal graphene flakes encapsulated by hexagonal boron nitride (hBN)[12] or in suspended graphene[13]. The doping method described here was applied to samples over 5 x 5 mm$^2$, resulting in graphene-based devices readily applicable to quantum resistance metrology, i.e. stable devices displaying quantum Hall effect at low magnetic fields (Fig. 1a). We show that air-stable molecular doping of graphene is achieved when organic molecules embedded in a polymer matrix, diffuse through the matrix, and spontaneously accumulate at the graphene surface due to formation of a charge-transfer complex. High electron affinity F4TCNQ molecules are incorporated in PMMA spin coated at ambient conditions on SiC/G, with graphene being used both as the target substrate for molecular assembly and, simultaneously, as a charge sensor. The stability of the samples allows us to study the chemical composition as well as electronic transport properties of the F4TCNQ/graphene system. Our devices include large (W=5 mm x L=5 mm) and small (W=2-50 μm x L=10-100 μm) epitaxial graphene Hall bars fabricated by electron beam lithography as described elsewhere[14]. After fabrication, devices were encapsulated by a 200 nm-thick, dopant-free PMMA layer to prevent drift in carrier concentration from ambient exposure and characterized by magnetotransport measurements.

The different chemical doping schemes of graphene devices are shown in Fig. 1b. When PMMA is used as a spacer between graphene and the molecular dopant layer, the carrier density decreases three orders of magnitude from its pristine value, n≈1x10$^{13}$ cm$^{-2}$ to near charge neutrality, n≈1x10$^{10}$ cm$^{-2}$ (Fig. 1c). Importantly, even at such low carrier densities the carrier mobility remains high, with the largest measured value exceeding μ≈50,000 cm$^2$/Vs at T=2 K in small devices. The effect of doping is homogeneous over millimeter scale and samples retain their low carrier density over the course of two years, even under ambient conditions (Supplementary S1). To achieve this, a 200 nm-thick layer of the polymer-F4TCNQ dopant blend is spin coated onto a PMMA-protected sample, followed by thermal annealing above the PMMA glass transition temperature. The resulting carrier density could be fine-tuned by the total annealing time. For a concentration of 7 wt.% of F4TCNQ in PMMA, charge neutral graphene is achieved after annealing at T=160 °C for 5 minutes. Shorter annealing times yield hole-doping and longer times yield electron-doping (Supplementary S1). Using the optimal time, we have consistently observed a decrease in electron density by three orders of magnitude together with a ten-fold increase of carrier mobility at T=2 K in more than 20 devices on 13 different substrates. Typical carrier concentrations in doped samples are of the order of $5 \times 10^{11}$ cm$^{-2}$ at room temperature. They decrease to values $<1 \times 10^{10}$ cm$^{-2}$ at T=2 K, with corresponding carrier mobilities in the range of 30,000-50,000 cm$^2$/Vs (Fig. 1d). The PMMA spacer layer plays a crucial role in achieving high carrier mobilities. While both PMMA and the dopant blend act independently as moderate p-dopants when deposited directly on SiC/G, it is only when the PMMA spacer layer is incorporated between graphene and the dopant blend that we observe the near to charge neutrality doping effect. Similarly, carrier mobilities exceed 10,000 cm$^2$/Vs only if the PMMA spacer and dopant blend operate in tandem.

We show that the doping of graphene is the result of F4TCNQ molecules diffusing through the PMMA layer and accumulating at the graphene surface. Figures 2a, b show the chemical depth-profile of the polymer stack, obtained by Time-of-Flight Secondary Ion Mass Spectrometry (ToF-SIMS), revealing both diffusion of F4TCNQ through the PMMA spacer and the accumulation of molecules at the graphene surface. From this we estimate the diffusion coefficient of F4TCNQ through PMMA to be of the order of $10^{-14}$ cm$^2$/s and by integrating the areas under the ion current intensity curves, we estimate the density of F4TCNQ near the graphene surface to be $\sim 4.6 \times 10^{14}$ molecules/cm$^2$ (Supplementary S2). Graphene and metallic surfaces promote the accumulation of F4TCNQ, while there are virtually no dopant molecules at the polymer/SiC interface. The surface density of F4TCNQ is roughly 50% greater on graphene (and 6-fold higher on gold) compared to that in the dopant blend layer (Fig. 2c). We attribute the accumulation of F4TCNQ on the graphene surface and the measured p-doping effect to the formation of a charge transfer complex, with partially charged F4TCNQ remaining at the graphene interface to preserve overall charge neutrality. F4TCNQ is known to be mobile in thin polymer films[15,16], with its diffusion depending on polarity and glass transition temperature of the polymer. When using an inert PMMA as a host matrix, F4TCNQ remains neutral both in the doping layer and as it diffuses through PMMA spacer layer[17]. The formation of a charge transfer complex takes place only when encountering an electron donor, such as graphene. Once charged, the F4TCNQ anion is bound to graphene, stabilized by Coulomb interaction[18].

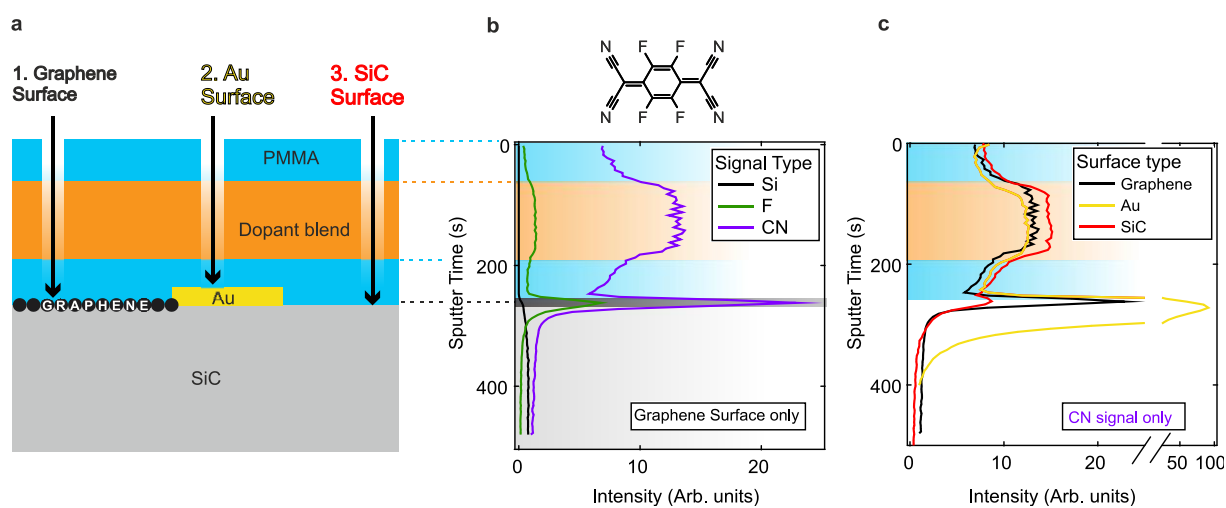

**Figure 2.** Chemical profiling of the polymer layers using ToF-SIMS to detect fingerprints of the F4TCNQ molecule (F and CN ions) as one probes deeper into the polymer stack. (a) Samples were prepared with PMMA spacer, dopant blend and PMMA encapsulation layer. Three distinct spots on the substrate have been investigated: 1. graphene, 2. gold and 3. SiC surfaces. (b) When analyzing the polymer layers on spot 1, above the graphene surface, the ion intensity for F and CN ions (top inset shows a schematic representation of a F4TCNQ molecule) versus sputter time reveals significant accumulation of F4TCNQ at the graphene/PMMA interface, as well as the spatial distribution of F4TCNQ molecules through the spacer and encapsulation PMMA layers. The onset of the silicon signal (Si) is the marker which indicates that the SiC substrate has been reached. The shaded regions denote the extent of each layer, from top PMMA layer down to SiC substrate. (c) Here we focus only on the CN signal measured at all three different spots. This analysis shows accumulation of F4TCNQ on the conductive surfaces of graphene and gold, but virtually no accumulation on SiC.

We investigated further electron transport details of the F4TCNQ-graphene charge-transfer complex system by introducing a top electrostatic gate (Fig. 3a, inset). The top gate enables additional fine tuning of the carrier concentration within $\Delta n \sim 5 \times 10^{11}$ cm$^{-2}$ using gate voltages $V_G$=-100 V to +200 V. At $V_G$=0 V the carrier density is n=$5 \times 10^{11}$ cm$^{-2}$ at room temperature and graphene is in the metallic limit. In this case, $\rho_{XX}$(T, $V_G$=0 V) decreases linearly with temperature from its room temperature value, due to suppression of acoustic phonon scattering. Quantum corrections to resistance result in a log(T) dependence below the Bloch-Grüneisen temperature $T_{BG} = 2v_{ph}E_F/(k_B v_F) = 2v_{ph}\hbar v_F \sqrt{\pi n}/(k_B v_F) \approx 38$ K, with the phonon velocity $v_{ph}$=$2 \times 10^4$ m/s ($E_F = \hbar v_F \sqrt{\pi n}$ the Fermi level of graphene, ℏ the reduced Planck's constant, $k_B$ the Boltzmann constant and $v_F$=$10^6$ m/s the Fermi velocity in graphene[19]). In contrast, when graphene is gated to the Dirac point, the sheet resistance $\rho_{xx}$(T, $V_g$=-50 V) monotonically increases as the temperature drops, though remains finite with no indication of a transport gap in the current voltage characteristics down to T=2 K (Fig 3a).

To characterize the magnitude of carrier density fluctuations (so-called charge puddles) we conducted low-temperature magnetotransport measurements on chemically and electrostatically gated devices and found these fluctuations to be on the level of $\Delta n \sim \pm 6 \times 10^9$ cm$^{-2}$ ($\Delta E_F \sim \pm 9$ meV). Fig. 3b shows longitudinal resistance versus carrier concentration in which every data point, extracted from individual measurements of $\rho_{XX}$(B), $R_{XY}$(B) at a fixed gate voltage, corresponds to devices behaving as a system with a single electronic band and spatially homogenous carrier density[20–25]. That is, data points in Fig. 3b fulfill simultaneously the criteria of linear $R_{XY}$(B) at low fields[20–23], and fully developed half-integer Quantum Hall Effect at high fields[24,25], i.e. $\rho_{xx}$(B)=0 Ω and strictly quantized $R_{XY}$ plateau over the entire available range of magnetic field (Supplementary S3). The gap in data around zero carrier concentration thus corresponds to data points where graphene is in the charge puddle regime.

Under quantizing conditions, the residual disorder in the sample causes non-zero and oscillatory $\rho_{xx}(B)$ once the magnetic length approaches the average charge puddle size[24,25]. In our samples with low carrier density concentration ~$10^{10}$ cm$^{-2}$ we have observed no deviation from $\rho_{xx}(B)=0$ to the largest magnetic field available in our setup (B=14 T) (Supplementary S4). Thus, we establish an upper limit for the puddle size of about $l_B = \sqrt{\hbar/eB} \approx 7$ nm. The charge puddle magnitude is directly connected to disorder introduced by, e.g. topography or inhomogeneous doping[26]. The small magnitude of charge puddles measured in our devices indicate that SiC/G doped with F4TCNQ molecules is homogenous also at the microscopic scale, with carrier density fluctuations comparable to those in high-quality, hBN-encapsulated graphene flakes[12] (Supplementary S4).

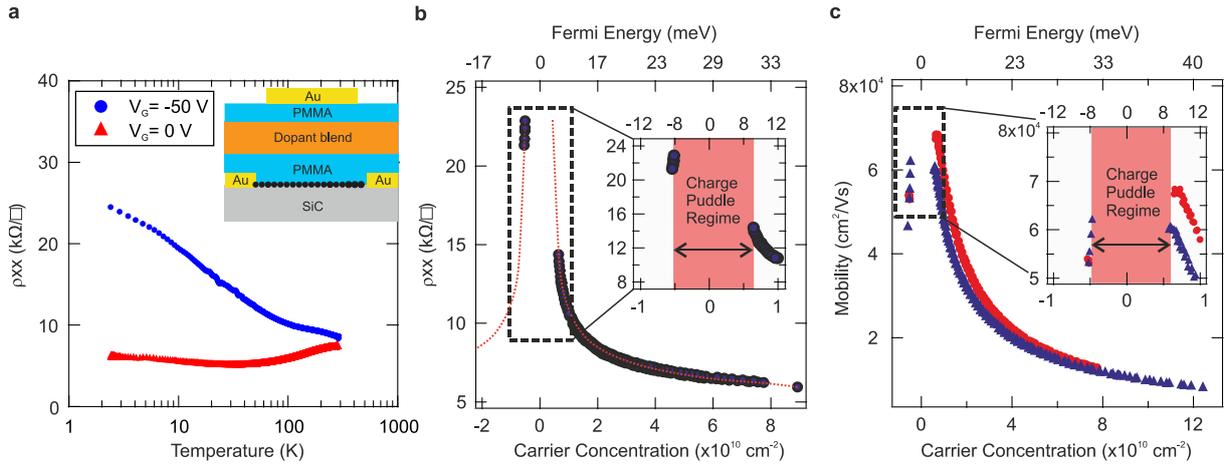

**Figure 3.** Electrostatic gating of chemically doped graphene. (a) Temperature dependence of longitudinal resistance for graphene in the metallic limit (red) and gated to Dirac point (blue) (Inset) Schematic representation of a chemically doped graphene device with a metallic gate (the topmost Au layer). (b) $\rho_{xx}$ versus carrier concentration shows the characteristic Dirac peak around the charge neutrality point (red dotted line serves as a guide to the eye). The Dirac point is crossed below Vg=-40 V, and the device has well-defined carrier densities within ±6x10$^9$ cm$^{-2}$ at T=2 K, which corresponds to a Fermi energy of ±9 meV (c) Corresponding charge carrier mobilities, with values up to 70,000 cm$^2$/Vs. Each point in (b) and (c) represents data of magnetic field scans where R$_{XY}$ is linear in the low magnetic field limit and the device shows fully developed Quantum Hall Effect at high magnetic fields ($\rho_{xx}=0$ and exactly quantized R$_{XY}$ plateaus). The gap in data around zero carrier concentration corresponds to omitted data points where graphene is in the charge-puddle regime.

It is remarkable that epitaxial graphene displays such high carrier mobilities and low disorder even at extreme dopant coverage, being decorated with a dense layer of molecules of about 3-4 molecules nm$^{-2}$ ($c \approx 3 \times 10^{14}$ molecules/cm$^2$, comparable to the molecular coverage of $c \approx 4.6 \times 10^{14}$ molecules/cm$^2$ from SIMS). We estimated the molecular density at the graphene surface from the shift in carrier density measured in doped graphene with respect to its pristine concentration ($\Delta n \approx 1 \times 10^{13}$ cm$^{-2}$), and assuming that 0.3 electrons are withdrawn from epitaxial graphene per molecule[27,28] with 1/10 gate efficiency[11]. The homogeneity in doping is in part enabled by the high degree of F4TCNQ dispersion inside the PMMA matrix, shown by room temperature grazing-incidence wide angle x-ray scattering (GIWAXs). The diffractogram in fig. 4a,b reveals a broad amorphous halo with a distinct diffraction peak at q=9.6 nm$^{-1}$ from PMMA (Supplementary S5). The absence of diffraction spots from F4TCNQ implies the lack of molecular aggregation (i.e. crystallites) inside the matrix. Given the size of the F4TCNQ molecule, we propose that at this packing density a feasible molecular orientation of F4TCNQ is close to that of molecules standing up on the graphene surface[28]. Yet, we do not rule out molecular re-orientation or thermally-induced redistribution of charges in the dopant layer under the effect of electric field, even at low temperatures. Such charge redistribution in the dopants in close vicinity to graphene may be responsible for screening charge inhomogeneities that facilitate highly uniform doping[29]. Thermally activated motion of charges in the dopant layer is a plausible source of the hysteresis in $\rho_{XX}(T)$ when devices are subjected to thermal excursion from T=2 K to 230 K (Fig. 4a). More accurate resistance measurements reveal that charges in the dopant layer are mobile down to T$\approx$113 K (Supplementary S6), in notable coincidence with the energy scale of the measured charge inhomogeneity in doped epitaxial graphene ($\Delta E_F \sim \pm 9$ meV).

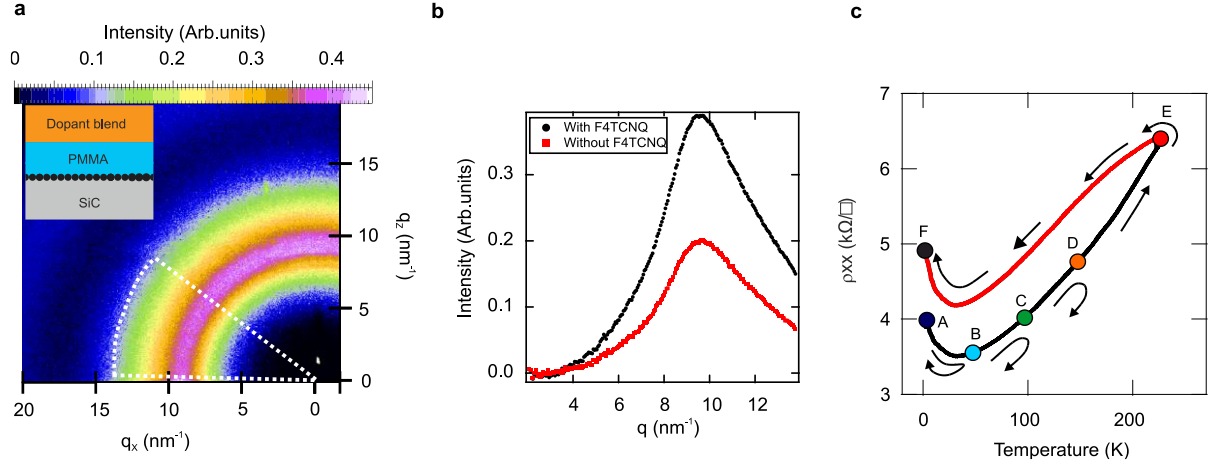

**Figure 4.** F4TCNQ in PMMA matrix. (a) Grazing incidence wide angle x-ray scattering (GIWAXs) measurements taken on SiC/G with dopant blend and PMMA spacer (inset), (ambient condition, room temperature, 0.15 degrees incident angle). The diffraction ring with radius q=9.6 nm$^{-1}$ from PMMA and the absence of F4TCNQ diffraction spots suggests a good molecular solubility in the polymer matrix. The color bar shows the normalized intensity, after a constant diffuse scattering background has been subtracted. The white dotted region denoted over which azimuthal angles the intensity profile in (b) has been averaged. (b) The reference sample, without F4TCNQ in the dopant blend displays a clear peak diffraction ring with radius q=9.6 nm$^{-1}$; the addition of F4TCNQ molecules enhances the signal two-fold. (c) Stability of doping at low temperature in a device that has been cooled down from room temperature. Starting at T=300 K we applied a gate voltage $V_G$=+50 V and kept down during cool down Once we reached T = 2 K, the gate terminal is set to $V_G$=0 V and sample sheet resistance acquires a value of $\rho_{XX}$=4 kΩ (point A). Thermal excursions to T=50 K (B), 100 K (C), 150 K (D) result in reversible $\rho_{XX}$(T) along the black curve. Once temperature exceeds T=150 K (E), $\rho_{XX}$(T) irreversibly changes to a higher resistive value (red curve), and on cooling down back to T=2 K, the sample resistance takes a value of $\rho_{XX}$ =5 kΩ (F) In the absence of the gate voltage, the sample resistance remains in the higher resistance $\rho_{XX}$(T) branch.

A possible explanation for the observed high charge carrier mobility of SiC/G at large molecular coverage is a high degree of spatial correlation between adsorbed F4TCNQ and impurities present on bare graphene. Together with the low charge carrier density fluctuations, increasing impurity (e.g. F4TCNQ) density can lead to suppression of charge scattering in graphene if there is sufficient inter-impurity correlation present in the system[30,31]. The actual organization and conformation of charged molecules on graphene is the result of delicate balance between molecule-graphene interactions as well as the intermolecular many body dispersion forces[32,33] in the presence of polymer. While the exact arrangement of molecules on graphene is difficult to probe through the thick polymer layers with surface science techniques, the graphene itself utilized as detector allows to gain an insight of such molecular reorganization at the polymer-graphene interface.

In summary, we presented a method of guiding the assembly of molecular dopants onto the surface of graphene by using an organic polymer matrix. With the doping stability observed in our samples, the tuneability of molecular coverage, and given the vast catalogue of polymers and organic/organo-metallic molecules, we expect this method to open up a scalable route towards expanding the properties and functionality of graphene and other 2D materials well beyond doping control. Moreover, the presented analysis of chemical and electron transport properties of doped graphene sheds light on the complex processes that molecular dopants undergo when embedded in polymers. This is relevant to the understanding of performance, materials, and interfaces in organic electronic devices, especially when combined with 2D materials. This method can be explored in the future to create and study electron transport properties of novel two-dimensional systems of ordered molecular arrays templated by 2D crystals[34,35].

**Acknowledgements.** We thank Dmitry Golubev, Alexander Tzalenchuk and Tord Claeson for useful discussion and critical reading of the manuscript. This work was jointly supported by the Swedish Foundation for Strategic Research (SSF) (No. IS14-0053, GMT14-0077, RMA15-0024), Knut and Alice Wallenberg Foundation, Chalmers Area of Advance NANO, the Swedish Research Council (VR) 2015- 03758,n the Swedish-Korean Basic Research Cooperative Program of the NRF (No. NRF-2017R1A2A1A18070721). We thank CHESS for providing time for GIWAXs measurements. CHESS is supported by the NSF & NIH/NIGMS via NSF award DMR-1332208.

**Data Availability.** The authors declare that the main data supporting the findings of this study are available within the article and its Supplementary Information files. Extra data are available from the corresponding author upon request.

## Author Contributions

H.H. and S.L.A. fabricated samples, performed transport measurements and analyzed data. K.H.K. A.D. D.M., F.L., Y.W.P. and T.B. performed transport experiments. P.M. performed and analyzed SIMS measurements. K.M.P., L.Y. and C.M. analyzed GIWAXS measurements. H.H. and S.L.A. wrote the manuscript, with contributions from all of the authors. S.K. and S.L.A. conceived and designed the experiment.